\documentclass[12pt]{article}
\usepackage[english]{babel}
\usepackage[dvips]{epsfig}

\def\gtorder{\mathrel{\raise.3ex\hbox{$>$}\mkern-14mu
 \lower0.6ex\hbox{$\sim$}}}
\def\ltorder{\mathrel{\raise.3ex\hbox{$<$}\mkern-14mu
 \lower0.6ex\hbox{$\sim$}}}

\begin{document}

\centerline{\bf \LARGE{Two--photon exchange and elastic scattering}}
\centerline{\bf  \LARGE{ of electrons/positrons on the proton. }}
\vspace{.4cm}
\centerline{\Large \it{ (Proposal for the experiment at VEPP-3.) }}
\vspace{1cm}

{\large  J. Arrington$^{a}$, V.F. Dmitriev$^{b}$,
R.J. Holt$^{a}$, D.M. Nikolenko$^{b}$,
I.A. Rachek$^{b}$, Yu.V.Shestakov$^{b}$, V.N. Stibunov$^{c}$, D.K. Toporkov$^{b}$,
H.de Vries$^{d}$.} \\ 
{\it 

$^{a}$ ANL, Argonne, IL 60439-4843, USA,

$^{b}$ BINP, 630090 Novosibirsk, Russia, 

$^{c}$ NPI at TPU, 634050 Tomsk, Russia,

$^{d}$ NIKHEF, P.O. Box 41882, 1009 DB Amsterdam, The Netherlands } \\ \\

\section {Introduction.}
The study of the electromagnetic form factors of the proton -- important 
properties of this fundamental object -- allows increased understanding of the nature of the 
proton, as well as the nature of interactions of its constituents - the 
quarks.
Until recently, the electric ($G_E(Q^2)$) and 
magnetic ($G_M(Q^2)$) form factors, which describe the distribution of charge and
current inside the proton, were determined by the separation of 
longitudinal and transversal contributions to the 
electron--proton scattering cross section.
Differential cross section of the elastic scattering in one--photon approximation,
assuming $P-$ and $T-$invariance, can be written \cite{rosen} as: 

\begin{eqnarray*}
\frac{d\sigma}{d\Omega} & = & \sigma_{Mott} \left[
\frac{G_E^2 + \tau G_M^2}{1+\tau} +2 \tau G_M^2 \tan^2
\frac{\theta}{2}\right],
\end{eqnarray*}
where $\sigma_{Mott}$ is the Mott cross section, $\theta$ is electron scattering
angle, $Q$ is transfered four-momentum, and $\tau = Q^2/4M_p^2$. 
Introducing the longitudinal virtual photon polarization, 
$\epsilon = (1 +2(1 + \tau)\tan^2(\theta/2))^{-1}$, one can re-write 
the above formula as:

\begin{eqnarray*}
\frac{d\sigma}{d\Omega} & = & \frac{ \tau\sigma_{Mott} }{\epsilon (1 +
 \tau)} \left[ G_M^2 + \frac{\epsilon}{\tau} G_E^2 \right] 
 \,\,\,.
 \end{eqnarray*}

The two form factors can be disentangled by measuring scattering cross sections
at different initial electron energies and scattering angles while keeping 
momentum transfer ($Q$) the same. Such a procedure is called Rosenbluth
separation or Rosenbluth technique. 
As is seen from the last formula, the contribution of the electric form factor
to the cross section drops down with increasing  $Q^2$. Therefore it becomes
difficult to measure $G_E$ using the Rosenbluth method at high $Q^2$.

\begin{figure}
\centerline{\epsfig{figure=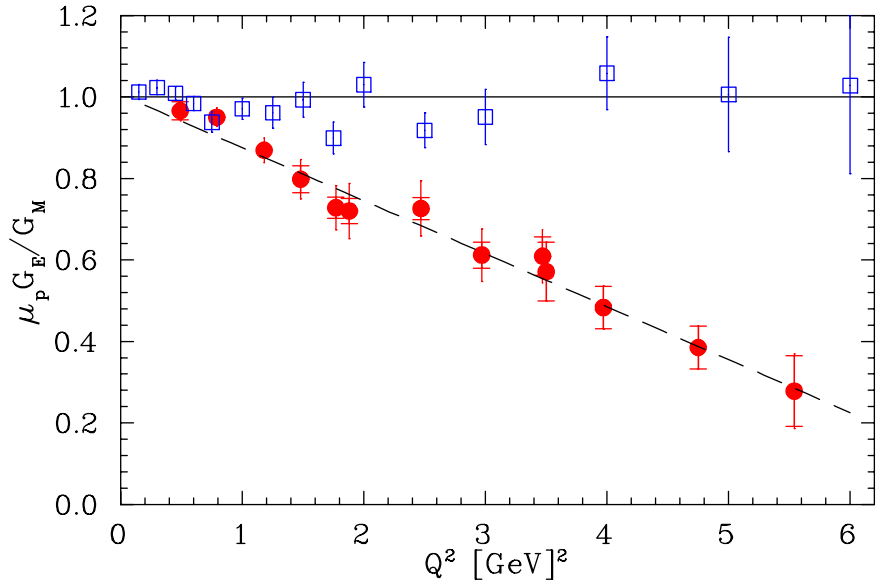, width=12cm}}
\caption[]{
From \cite{arr1}: comparison of form factors ratio, obtained by
Rosenbluth technique (hollow squares) with data of polarized measurements
(full circles).
}
\label{arrin1}
\end{figure}

In the mid-nineties, it became possible to use polarization transfer experiments 
to study nucleon electromagnetic form factors. Through such measurements 
contributions of interference terms into the scattering cross-section 
become accessible, hence the contribution of the small form factors can be
enhanced resulting in increased accuracy of their determination.
A series of precise measurements of the ratio of proton
form factors $G_E(Q^2)/G_M(Q^2)$ for a wide range of transfered momentum
was carried out recently at TJNAF~\cite{jlab}.
In these experiments a ratio of transverse ($P_t$) and longitudinal ($P_l$) 
polarization of recoil protons from elastic scattering of longitudinally
polarized electrons on an unpolarized hydrogen target was measured.
In such a case the ratio of proton form factors can be extracted 
directly from the ratio of $P_t$ and $P_l$:  
\begin{eqnarray*}
\frac{ G_E }{G_M} & = & -  
\frac{P_t}{P_l} \frac{(E + E')}{2 M_p}\tan \frac{\theta}{2} 
\,\,\,, 
 \end{eqnarray*}
where $E, E'$ are the electron energy before and after scattering and $M_p$ 
is the proton mass.
These polarization experiments yielded unexpected results, indicating
that the ratio $\mu G_E /G_M$ depends strongly on $Q^2$,
while before it had been assumed to be nearly constant, close to unity
(see Fig.~\ref{arrin1}).
A thorough reanalysis of the available unpolarized experimental
data \cite{arr1}, as well as new precise unpolarized measurements done at 
TJNAF~\cite{supros} have clearly shown that these two methods deliver contradictory
results.

\begin{figure}
\centerline{\epsfig{figure=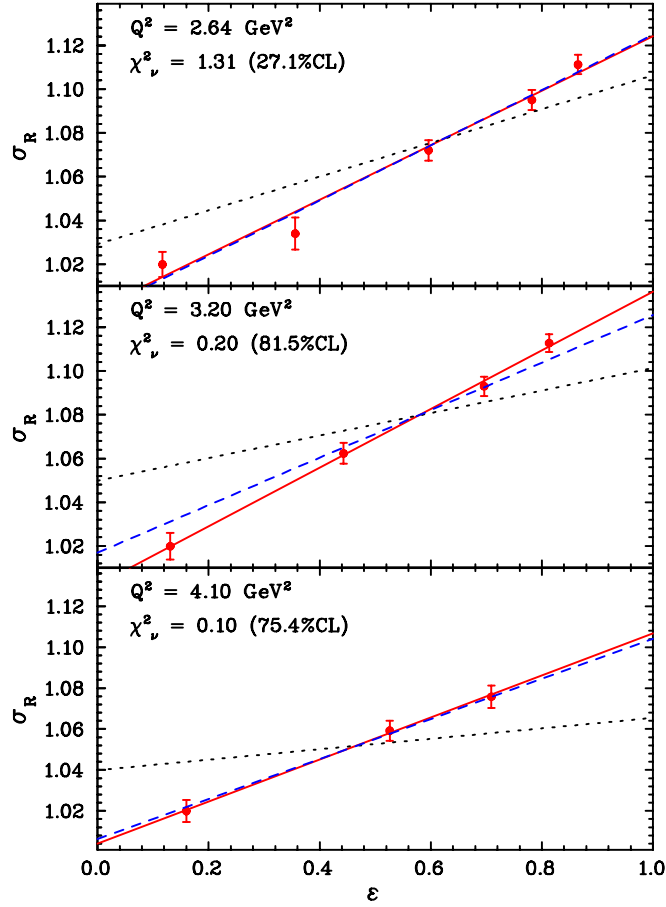,width=9cm}}
\caption[]{
From \cite{supros}: comparison of the results of a new measurement of the proton
form factors (``reduced'' cross section $\sigma_R \propto \tau G_M^2 + \epsilon G_E^2$
is presented, scaled to yield $\sigma_R \approx 1$ at $\varepsilon=0$) 
obtained by the Rosenbluth technique (points -- data, 
solid lines -- best linear fit)
with the world ``unpolarized'' data (dashed lines are linear fit to these data)
and with the ``polarized'' results (dotted lines show best liner fit to these data).
One can see that old and new ``unpolarized'' data are compatible, while they both
contradict the polarization measurements.
}
\label{supr}
\end{figure}

\section {Two--photon exchange.}
\begin{figure}
\centerline{\epsfig{figure=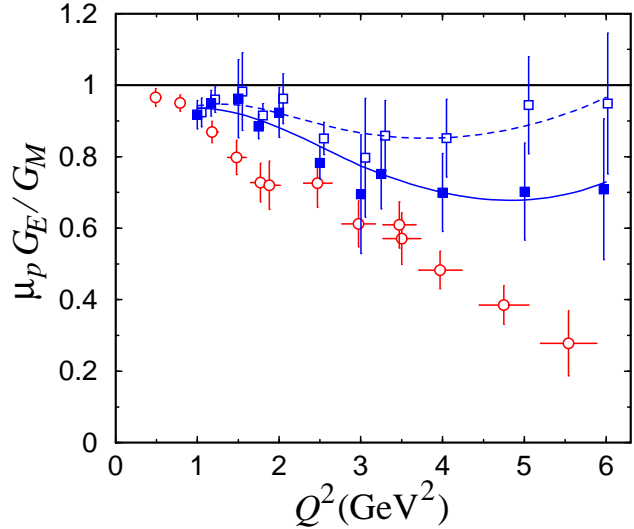, width=9cm}}
\caption[]{
From \cite{blunden}: The ratio of proton form factors obtained by the
Rosenbluth technique: before (hollow squares) and after (full squares) 
the TPE corrections, and the data of polarized measurements (hollow circles).
}
\label{blun}
\end{figure}

A number of authors \cite{rekal} - \cite{chen} argue  % , guich, arr2, blunden, afan
that the possible origin of these discrepancies is the failure of the 
one--photon approximation to precisely describe the results of unpolarized experiments. 
Indeed, with increasing  $Q^2$ the cross section of elastic scattering and
especially the contribution of electric form factor drops down substantially.
In such a case the contribution of two--photon exchange (TPE), which
depends weakly on  $Q^2$, can become considerable.
Thus it was shown in \cite{blunden} that allowing for the TPE in 
Rosenbluth technique, using a simplified model (which does not
include nucleon excitations in the intermediate state), leads to a substantial
change of form factors ratio, despite the small contribution of the TPE to 
the cross section (Fig.~\ref{blun}).
This results from a strong dependence of the TPE contribution on electron
scattering angle for a fixed $Q^2$ \cite{blunden}.

Complications arising in the calculation of the TPE corrections 
are connected with difficulties in accounting for proton excitations in the 
intermediate state.  The intermediate state contributions have been treated
in a recent calculation at the quark-parton level, using generalized parton
distributions to ~\cite{chen}.  

The Born amplitude is proportional to the lepton charge, $e_l$, while the TPE
amplitude is proportional to $e_l^2$.  The Born cross section is proportional
to $e_l^2$, while the interference term to the
cross section goes like $e_l^3$.  Hence the interference term, which is the dominant
part of the TPE contribution (since the TPE amplitude is small compared to the
Born amplitude) changes sign with respect to the Born cross section and can therefore
be determined by comparing electron--proton and positron--proton 
scattering.
\begin{figure}
\centerline{\epsfig{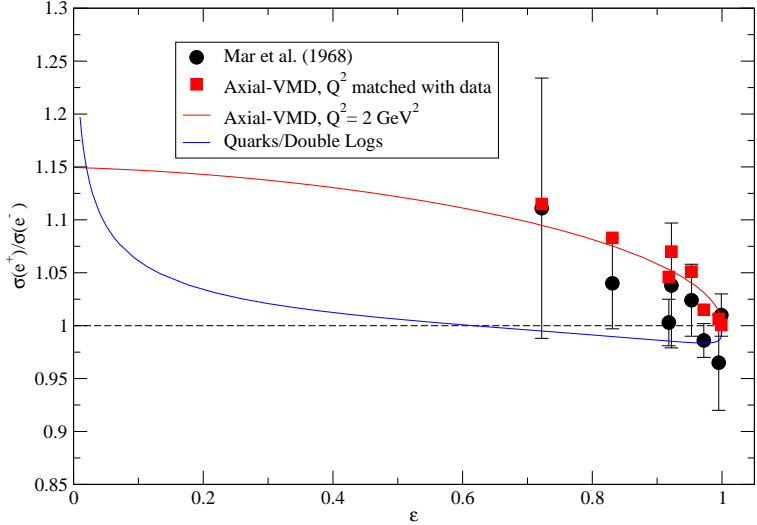}}
\caption[]{
From \cite{jlabl}: theoretical predictions from \cite{afana} 
(square points and curves) and
experimental data from \cite{mar} (circles) for the ratio 
$R = \sigma(e^+)/\sigma(e^-)$ as a function of $\epsilon$.
}
\label{jlbl}
\end{figure}

Attempts to measure the TPE contribution were made in the 1960s, but either 
the accuracy of the measurements was insufficient:  
$ \delta R/R \sim 5$\%,
where $R = \sigma(e^+)/\sigma(e^-)$, or scattering angles were too small
and therefore $\epsilon \approx 1$ -- where most theories predict 
$R \approx 1$,  see Fig.\ref{jlbl}.

\section {Experiment at VEPP--3.}
We propose to perform a measurement of $R$ at the VEPP--3 storage ring at an
energy of electron/positron beams of 1.6 GeV and at electron/positron
scattering angles approximately $25^o$, $65^o$ (corresponding to
$\epsilon \approx  0.90$, $0.45$ and $Q^2 \approx  0.3$, $1.5$ $GeV^2/c^2$)

There are several necessary preconditions for a successful realization of this
experiment: 
\begin{itemize}
\item 
the injector delivering positrons at a rate 50 $\mu$A/s (measured in 
VEPP--3 current);
\item 
the storage ring where electrons or positrons move in the same direction;
\item 
the experimental straight section with quadrupole lenses (to compress the
beam) and with an internal gas target; 
\item 
mean experimental luminosity (defined by positrons) can be rather high:
$L = I \cdot t = 0.009 \cdot 6 \cdot 10^{18} \cdot 10^{15} = 5 \cdot 10^{31}$,
here $t$ -- target thickness, $I$ -- positron current averaged over a working
cycle which includes storage phase, energy ramping phase, production phase
and return-to-storage-mode phase.
Storage phase for electrons is much less, as a result the cycle length is
about a factor 1.6 smaller than that for positrons and the luminosity is
larger by the same factor.
\item 
the equipment used at the previous experiment -- particle detectors, data 
acquisition system, detector and target infrastructure, readout and slow control
software as well as a large experience of conducting  internal target
experiments at VEPP--3 provides with a good basis for developing new target
and detector and performing the measurements.
\end{itemize}

\section {Target.}
In previous experiments at VEPP--3 a polarized deuterium target was used.
The luminosity was restricted by maximum achievable polarized target thickness
of $ t \approx 10^{14}$at/cm$^2$,  
while electron beam current was limited (by large current effects) at a level
of $I_{max}\approx$ 140 mA.
Mean luminosity was $L \approx 5 \cdot 10^{31}$.

In the new target we are going to utilize a similar storage cell: having
elliptical cross section  13$\times$24 mm, length 400 mm, cooled by 
liquid nitrogen.
Hydrogen flux directed to the cell is going to be restricted at a level of
$\sim10^{18}$ at/sec, providing a target thickness of $\sim10^{15}$at/cm$^2$.
Note that gas density profile along the cell has a triangle shape with
a maximum $\sim 5\times 10^{13}$at/cm$^3$ in the cell center 
and a base equal to the cell length (40 cm).
With these target parameters two upper limits are reached simultaneously:
first, with a higher gas flux the available vacuum pumps might not be able 
to maintain a required vacuum in the storage ring, second, for a thicker
target the detector singles rates might become too high.

Let us also note that e.g. doubling the target thickness would result in
only 30\% increase of an average luminosity, because it is the time of
positrons storing that takes up the main part of experimental time--cycle.

Gas flux to the storage ring vacuum chamber can be made about factor two
lower (while retaining target thickness the same) by decreasing the temperature
of the cell (down to $\sim 20^o K$). This can be done using an appropriate
cooler, but at present it is not available and has to be purchased.

A storage cell placed in the VEPP--3 ring would prevent beam injection 
into VEPP--4 and this may become a serious obstacle. That is why we are
going to modify the design of the coupling between the cell and the straight
section
in order to be able to remove easily the cell from the aperture of VEPP--3
during VEPP--4 operations.  

Optimal relation between beam storage time and data taking time in a 
timing cycle for positrons is shown in Table~\ref{tbl}. Positron beam 
current will be changing from 50 mA down to 9 mA during the data taking
phase. 
To decrease the systematic errors we are going to keep electron beam in the
same range.
But since the timing cycle for electrons is substantially shorter we will
run two ``electron'' cycles for each ``positron'' cycle.

\begin{table}[htb]
\caption[]{
Timing scheme for positron/electron working cycles (in seconds).

}
\begin{center}
\begin{tabular}{p{7cm}rr} \hline \hline
 phase    & positrons  & electrons    \\ \hline \hline
 storage   & 1630     &  10 \\ \hline
 energy ramping    &  300     &  300 \\ \hline
 production  &  1920   &  1920 \\ \hline
 return to storage mode&  300   &  300 \\ \hline \hline
sum for a single beam cycle & 4150   &  2530 \\ \hline \hline \hline
duration of a total cycle, %&   &   \\ 
which includes  %&   &   \\ 
3 beam cycles: $\{e^-/ e^+/  e^-\}$ & \multicolumn{2}{c}{~~9190}   \\ \hline \hline
\end{tabular}
\end{center}
\label{tbl}
\end{table}

\section {Detector, event rate.}
As was mentioned above, the detector for the measurement of ($e^+p$) and ($e^-p$) elastic
scattering will be build on the basis of the detector used in the experiment 
that measured the deuteron form factors.
Scattered electron and recoil proton will be detected in coincidence,
which allows us to use kinematical correlations between their emission angles and 
energies, characteristic of two--body reactions. 
This is important for separation of the events from the process 
under study from those of various background processes. 
  \begin{figure}
\centerline{\epsfig{figure=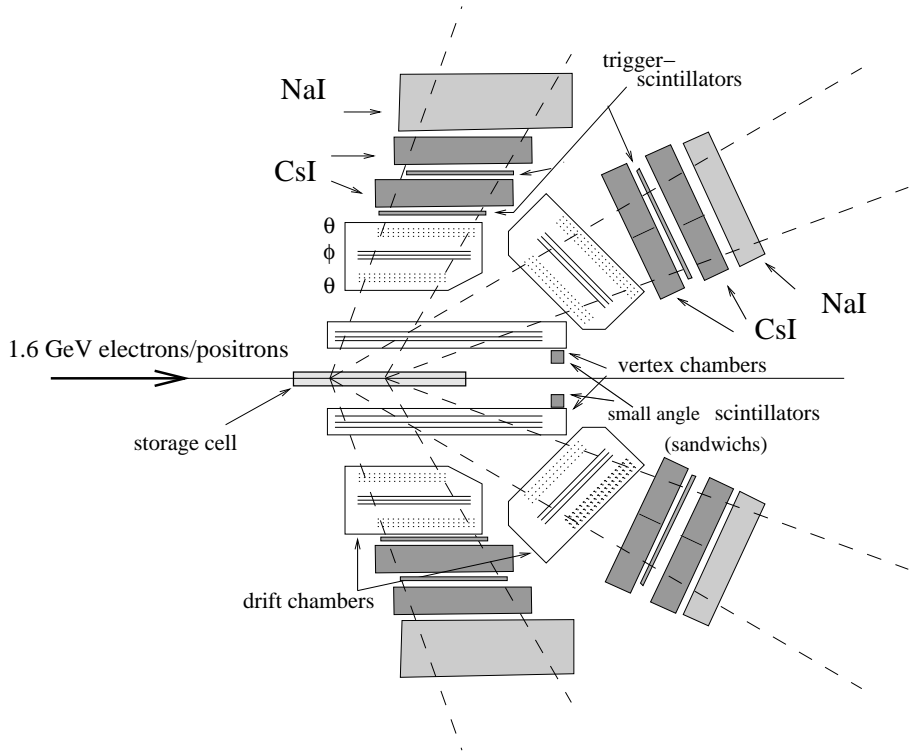, width=12cm}}
\caption[]{
Schematic side view of the particle detection system.
}
\label{det}
\end{figure}
 
The detector (Fig.~\ref{det}) is comprised of two identical systems placed 
symmetrically with respect to the storage ring median plane. 
Azimuthal angle acceptance of each system is 60$^o$.
Total $\phi$--acceptance is $\Delta\phi$ = 120$^o$, because particles can
be detected either by upper or by lower system.
Regarding the polar angles -- electrons/positrons scattered at angles
close to $12^\circ$, $25^\circ$, $65^\circ$ will be detected. These
three $\theta$--ranges are denoted later on as Small Angles ({\it SA}), 
Medium Angles ({\it MA}) and
Large Angles ({\it LA}).

Electrons/positrons scattered at {\it SA} are detected by small scintillators
(sandwiches). Recoil protons in this case are detected by {\it LA}--arms, where 
their trajectories and energies are measured.
Here only part of extended target is ``visible'' to the detectors.
Scattering at {\it SA} will be used as a luminosity monitor --- as it was
pointed out at small $Q^2$ and $\theta_e$ the ratio $R$ should be very 
close to unity, i.e.  $\sigma(e^+) \approx \sigma(e^-)$.
Application of two detector systems (upper and lower) not only increases 
the detecting solid angle but also allows to suppress systematic errors
related to instability of the electron/positron beam position.

When electrons/positrons scatter to {\it MA}--arms, protons hit 
{\it LA}--arm detectors. For these events tracks and energies are
measured for both particles. 
The experience obtained at previous experiments allows us to be sure that
the information gathered by such detectors will be quite sufficient for a
reliable event identification for scattering both at {\it SA} and at {\it MA}.  

In addition, we plan to equip the {\it LA}--arms with electromagnetic
calorimeters (there were no calorimeters here earlier).
This is needed in order to measure the energy of electrons/positrons 
scattered at those angles, to separate the elastic electron/positron from
pions or inelastic events.
Note that in this case recoil protons are detected by {\it MA}--arm detectors.
A minimal configuration of the calorimeters that can be assembled using 
CsI and NaI crystals, which we have at our disposal, is shown in Fig.~\ref{det}. 
 
 \begin{figure}
\centerline{\epsfig{figure=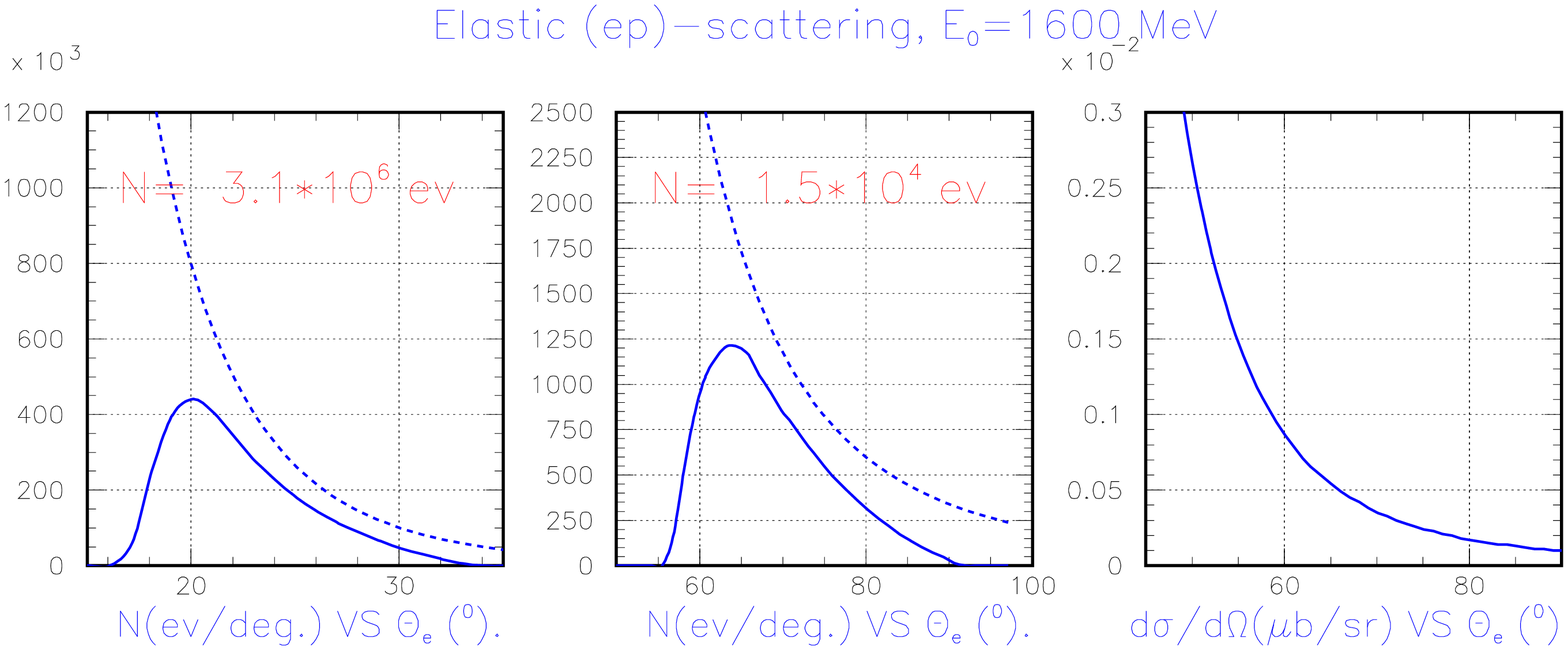, width=18cm}}
\caption[]{ 
Left and middle panels: solid lines show number of events (in a 
$\Delta \theta_e=1^o$  bins) detected by {\it MA}--arm (left panel) and
by {\it LA}--arm (middle panel) as a function of $\theta_e$ for a luminosity
integral $\int Ldt = 10^{15}$at/cm$^2 \cdot$12kQ (or during 30 days run).
Total number of events detected by {\it MA}--arm is $3.1\cdot10^6$ and by
{\it LA}--arm is $1.5\cdot10^4$.
Dashed curves demonstrate the event counts that would be in case of 
a point--like target. 
The right panel shows differential cross section of the elastic 
($ep$)--scattering.
}
\label{even}
\end{figure}

\begin{figure}
\centerline{\epsfig{figure=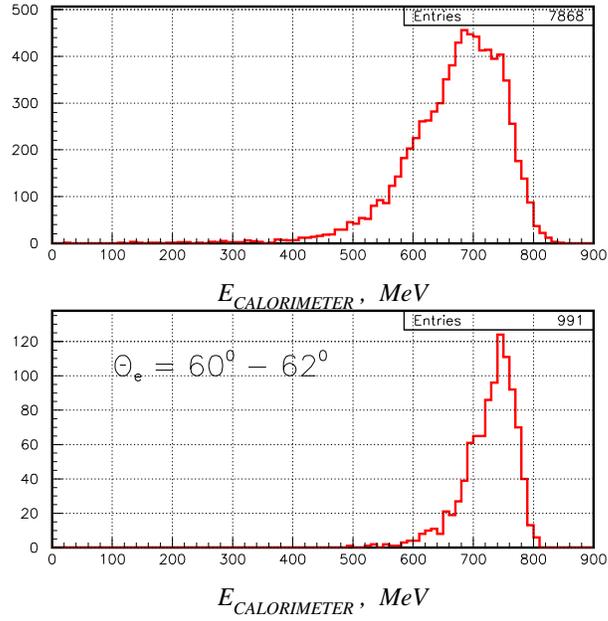, width=9cm}}
\caption[]{
Electron energy spectra measured by {\it LA}--arm calorimeter 
(Monte Carlo simulation) for events with the proton detected by 
{\it MA}--arm.
Upper histogram -- all events,  lower histogram -- for 
quasi-monochromatic electrons ($\theta_e=60-62^o $).
}
\label{electr}
\end{figure}

Cross section of ($e^+p$),($e^-p$) elastic scattering at {\it LA} is
two orders of magnitude lower than that for {\it MA}.
However, as it shown below, selection of the elastic scattering events
can be done in this case as well, by measuring trajectories of both 
particles and electron energy at an accuracy provided by the calorimeters
($\sigma \approx50$ MeV,  see Fig.~\ref{electr}).

\section {
Background estimation.
}
As it was mentioned above, difficulties in event selection may come out
only for scattering at {\it LA}, where the cross section of elastic scattering
becomes small   ($\sim 0.2-1$ nb/sr, see Fig.~\ref{even}).
Here we present an estimation of a background level from the processes 
which seem to be the most dangerous, namely 
electro- and photoproduction of pions.

Differential cross section of pion electroproduction can be written as
\cite{foster}:

\begin{eqnarray*}
\frac{d\sigma}{dE' \,d\Omega_e \,d\Omega_\pi} & = & \Gamma \frac{d\sigma}{d\Omega_\pi}\,\,\,,
\end{eqnarray*}
where $\Gamma$ is a flux of virtual photons, defined as:

\begin{eqnarray*}
\Gamma & = &  \frac{\alpha}{2\pi^2} \frac{E'}{E} \frac{W^2-M_p^2}{2M_pQ^2} \frac{1}{1-\epsilon}\,\,\,.
\end{eqnarray*}
Here $W$ is an invariant mass of the hadron system. 

\begin{figure}
\centerline{\epsfig{figure=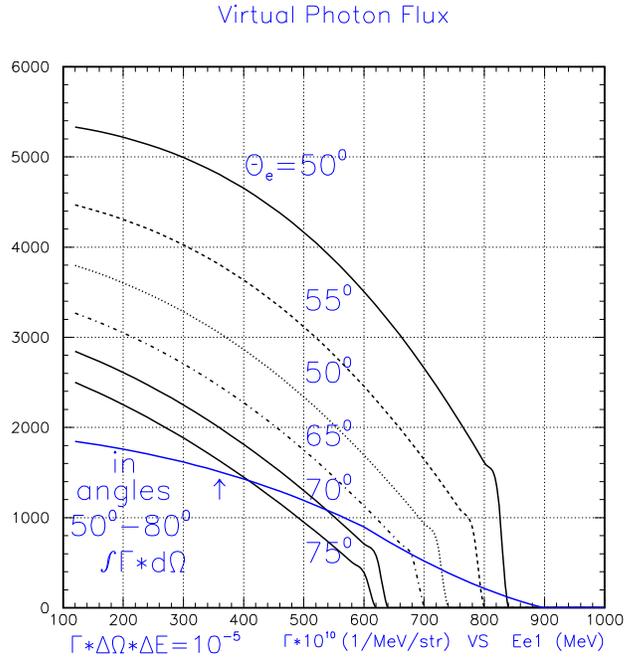, width=9cm}}
\caption[]{
Virtual photon flux $\Gamma$ as a function of energy of scattered electrons
for various $\theta_e$ (specified in the figure).
A gently sloped curve shows an integral of $\Gamma$ over solid angle of the
{\it LA}--arm.
}
\label{virt}
\end{figure}
From Fig~\ref{virt} one can see that 
after applying even a loose cut on scattered electron energy 
(e.g. $E'>400$ MeV) 
a total flux of virtual photons would not exceed $10^{-5}$ of 
a number of electrons deposited on the target. 
The cross section $d\sigma / d\Omega_{\pi}$ of the reaction 
$e p \rightarrow e \pi^+ n$ is estimated to be $\approx 10 \mu b/sr$, using
data from \cite{fro, egin}. The cross section of the reaction 
$e p \rightarrow e \pi^o p$ is about the same.
Therefore  $\Gamma \cdot d\sigma / d\Omega_{\pi}$ should not exceed 
$0.1 nb/sr$.
This is close to elastic scattering cross section, but applying cuts on the angular
correlations specific to elastic scattering allows one to further decrease 
this background by two orders of magnitude.

Electron scattering at forward angle with pion production is usually considered
as a process of pion photoproduction by equivalent photons.
In this case there is only one charged particle (besides the undetected 
electron),
therefore in order to get hits in tracking systems of both arms (from e.g
$\gamma p \rightarrow \pi^o p$ reaction) it is required that a gamma-quantum 
from
pion decay to be converted to a charged particle in materials on its way
to wire chambers.
Probability of such conversion ($K_{con}$) is small, estimated as $\sim10^{-2}$.
\begin{figure}
\centerline{\epsfig{figure=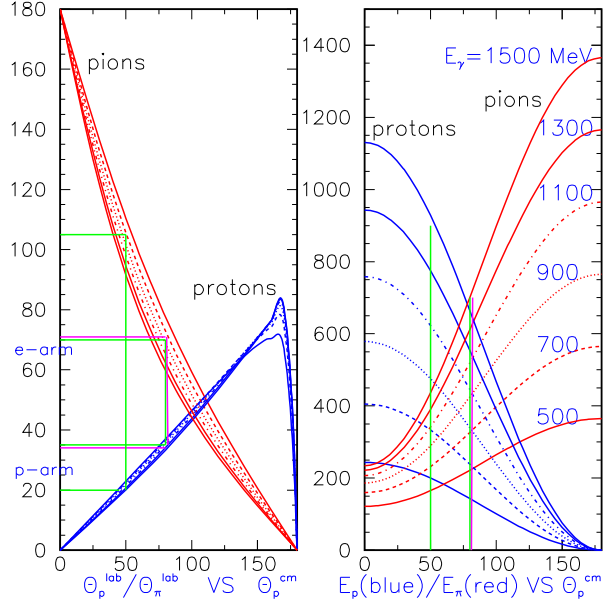, width=9cm}}
\caption[]{
Pion production kinematics.
Left panel: emission angles of protons (blue curves) and pions
(red curves) in Lab system as a function of proton C.M.--angle for various
photon energies (legend for curves is shown in the right panel). 
Right panel: kinetic energy of protons (blue curves) and pions (red curves)
in Lab system as a function of proton C.M.--angle for various
photon energies.
}
\label{kpi}
\end{figure}

As one can see in Fig.~\ref{kpi}, when a proton hits the {\it MA}--arm 
a $\pi^o$ is emitted in the direction of {\it LA}--arm and can be detected,
providing the conversion occurred. 
A number of equivalent photons emitted by a beam's electron 
(in energy intervals $\Delta E_\gamma$   $\pm$100 MeV) 
is shown in Table~\ref{tbpi} (second column)
for various photon energies (first column).
Cross section of pion photoproduction does not exceed $5\mu b/sr$ \cite{mainz}.
Taking into account small $K_{con}$ one gets an event rate about the same as
for elastic scattering (the last column in Table~\ref{tbpi}). However here
again the background can be greatly reduced by applying angular correlation
cuts.   

It is also worth noting that a requirement to have a large energy deposition in 
{\it LA}--arm calorimeter is important to avoid confusing
large--angle and medium--angle scattering events.

\begin{table}[htb]
\caption[]{ 
Estimation of neutral pion photoproduction background.
}
\begin{center}
\begin{tabular}{lcccc} \hline \hline
$E_{\gamma}$ &  $N_{\gamma}^{eq}$&$K_{con}$&
$d\sigma_{\pi} /d\Omega$ ($\mu$b/sr)&$
N_{\gamma}^{eq} \cdot K_{\gamma}^{con}\cdot d\sigma/d\Omega$ (nb/sr)\\ 
\hline \hline
500    & 0.0058   & 0.01   &5   &0.29  \\ \hline
700    & 0.0034   & 0.01   &5   &0.17   \\ \hline
900    & 0.0022   & 0.01   &5   &0.11   \\ \hline
1100   & 0.0016   & 0.01   &5   &0.08   \\ \hline
1300   & 0.0012   & 0.01   &5   &0.06   \\ \hline
1500   & 0.0010   & 0.01   &5   &0.05   \\ \hline
\end{tabular}
\end{center}
\label{tbpi}
\end{table}

\section {
Experiment duration. Estimation of accuracy.
}

The experiment is divided into three phases, each requiring either VEPP--3
hall access or VEPP--3 operation:
\begin{enumerate}
\item 
during shutdown of the VEPP--3 the following should be completed:
installation of the straight section with gas target, obtaining the vacuum,
commissioning the target, assembling and commissioning detectors,
data acquisition system and target control system ;  
\item 
beam tuning: finding/restoring regimes of positron/electrons beam 
operation at VEPP--3, minimizing background --- 7 days in total;
\item 
data taking run: to obtain high statistical accuracy in measuring $R$ 
(see Table~\ref{stat})
it is required approximately 300 full time-cycles (as defined 
in Table~\ref{tbl}) corresponding to $\approx$30 days of continuous
operation of the storage ring. 
\end{enumerate}

Expected statistical accuracy in measuring $R$ is shown in Table~\ref{stat}.
Below several possible sources of systematic errors are listed.
\begin{itemize}
\item
Unequal beam energy for electrons and positrons.\\
It is assumed that this difference can be made 1 MeV or less.
Then $\Delta R$/$R$ for three angle intervals will not exceed 
$0.1$\%, $0.2$\%, $0.2$\%.
Using {\it SA}--arm data for normalization one can reduce this error
by factor two. 

\item
Different beam positions for electrons and positrons.\\
Sensitivity of $\Delta R$/$R$ is estimated as $5.6$\%/mm, $1.4$\%/mm,
$0.9$\%/mm for {\it SA-, MA-, LA-}scattering respectively. 
Beam positions will be monitored by current pick--up sensors
with accuracy $\sim 0.1$mm for relative position of electron/positron beams.
Moreover already existing system for beam position
stabilization  can keep the beams orbit near the experimental straight section
sable with same accuracy. 
Besides, one can see that {\it SA}--arms are very sensitive to the beam 
position, hence their count rates can be served as a beam position monitor.  
Finally since we have symmetrical arm pairs therefore by averaging count 
rates of up-arm and down-arm this effect can be suppressed in first order.   
Here an accuracy of $\Delta R$/$R \sim 0.1$\% can be achieved.
\item
Time instability of detectors efficiency.\\
This effect (like the previous one) is suppressed in first order
because data collection with electron and positron beams will be alternated
regularly. 
We expect that detectors efficiency would not change by more than $\sim$ 1\%
during one time cycle. Assuming that this instability has a random character,
one would get a reduction of this error by averaging over many cycles.
For 300 time-cycles the reduction will be $1/ \sqrt{300}$, i.e. by over an
order of magnitude.
\item 
Drift of the target thickness in time. Special efforts will be devoted to
achieve a stable gas flow during the whole experiment. And again using 
data rate of {\it SA}--arms as a monitor, and averaging over many cycles 
one should get a contribution of this effect to the systematic error of $R$
to be 0.1\% or less.
\end{itemize}

\noindent
Combining the above uncertainties, the total systematic error for the largest $Q^2$ bin
is expected to be constrained below $\Delta R$/$R \sim 0.3$\%. 
Therefore the measurement accuracy will be defined
by statistical error.

\begin{table}[htb]
\caption[]{ 
Expected accuracy of the measurement of the ratio
$R = \sigma(e^+)$ / $\sigma(e^-)$. 
It is assumed that $N_-\,\,=\,\,2N_+$. 
}
\begin{center}
\begin{tabular}{lcccccc} \hline \hline
$\theta_e $ ($^o$) &  $Q^2$ (GeV$^2$) &  $\epsilon $  &  $N_+$ (events)  &  $\Delta R/R$ \% (stat) & $\Delta R/R$ \% (sys)  \\ \hline \hline
10--12              &0.08--0.11        &  0.98--0.98   &  $8.7\cdot10^6$ &     ---  &  0.30 \\ \hline
19--27              &0.26--0.47        &  0.94--0.88   &  $3.1\cdot10^6$ &     0.07 &  0.30 \\ \hline
60--80              &1.40--1.76        &  0.51--0.32   &  $1.5\cdot10^4$ &     1.00 &  0.30 \\ \hline
\end{tabular}
\end{center}
\label{stat}
\end{table}

\begin{figure}
\centerline{\epsfig{figure=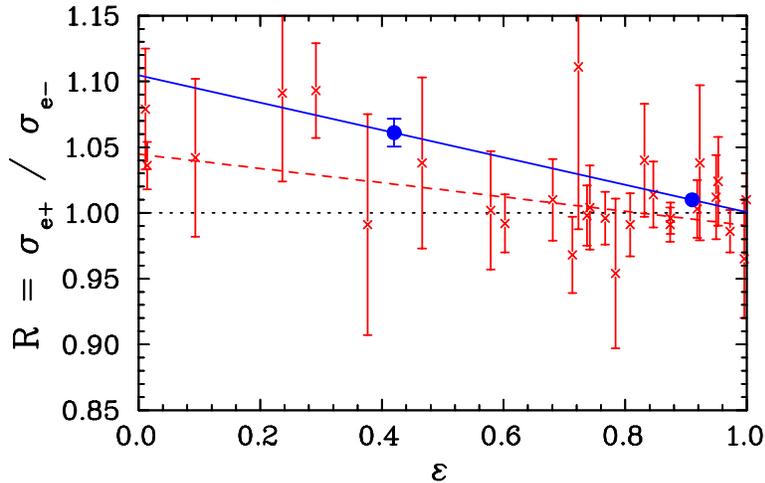,width=10cm}}
\caption[]{
Projected uncertainty (combined statistical and systematic) for the proposed measurement
(blue circles), compared to previous data (red ``x'' - Ref.~\cite{arr2}
and refs. therein).  Note that the previous measurements have an average
$Q^2$ value of approximately 0.5~GeV$^2$ for the data below $\epsilon=0.5$,
and thus should have a smaller TPE contribution than the proposed measurement.
The dashed line is a linear fit to the combined worlds data on $R$, and yields
a slope of $-(5.7\pm1.8)$\%
}
\label{proj}
\end{figure}

Calculations of the TPE effects indicate that the $Q^2$-dependence of the
TPE correction is weak at large $Q^2$ values.  Analyses of the discrepancy
between Rosenbluth and polarization transfer form factors also indicate a
weak $Q^2$-dependence, and a change in the $\epsilon$-dependence of 5--7\%
for $Q^2 > $1--2~GeV$^2$~\cite{arr1,guich,arr3}. Figure~\ref{proj} shows
the projected ratio and uncertainty, assuming an $\epsilon$-dependence
of roughly 5\% in the electron cross section, yielding a slope of $\approx$10\%
in the ratio of positron to electron yield.

This work is supported in part by the U.S. Department of Energy,
Office of Nuclear Physics, under contract W-31-109-ENG-38

\end{document}